\documentclass[aps, showpacs, showkeys, twocolumn]{revtex4-1}

\usepackage{amsmath,amssymb,amsthm,amsfonts,latexsym}
\usepackage{bm,amsfonts, mathtools}
\usepackage{graphicx,color, wasysym}
\usepackage{textcomp}
\usepackage{amsmath,amssymb,latexsym,epsfig}
\usepackage{appendix}

\def\ulamek#1#2{\mbox{\normalfont$\frac{#1}{#2}$}}

\DeclareMathOperator{\okr}{{\stackrel{{\scriptscriptstyle{\mathsf{def}}}}{=}}}
\DeclareMathOperator{\pokr}{{\stackrel{{\scriptscriptstyle{\mathsf{iii)}}}}{=}}}
\DeclareMathOperator{\D}{d\!}
\DeclareMathOperator{\E}{e} 
\DeclareMathOperator{\I}{i}

\begin{document}

\title[Composition law for the Cole-Cole relaxation and ensuing evolution equations]{Composition law for the Cole-Cole relaxation and ensuing evolution equations}

\author{K.~G\'{o}rska} 
\email{katarzyna.gorska@ifj.edu.pl}

\author{A.~Horzela}
\email{andrzej.horzela@ifj.edu.pl}

\author{A.~Lattanzi}
\email{ambra.lattanzi@ifj.edu.pl}
\affiliation{H. Niewodnicza\'{n}ski Institute of Nuclear Physics, Polish Academy of Sciences, ul. Eliasza-Radzikowskiego 152, PL 31342 Krak\'{o}w, Poland}

\begin{abstract}
Physically natural assumption says that the any relaxation process taking place in the time interval $[t_{0},t_{2}]$, $t_{2} > t_{0}\ge 0$ may be represented as a composition of processes  taking place during time intervals  $[t_{0}, t_{1}]$ and $[t_{1},t_{2}]$ where $t_{1}$ is an arbitrary instant of time such that $t_{0} \leq t_{1} \leq t_{2}$. For the Debye relaxation such a composition is realized by usual multiplication which claim is not valid any longer for more advanced models of relaxation processes. We investigate the composition law required to be satisfied by the Cole-Cole relaxation and find its explicit form given by an integro-differential relation playing the role of the time evolution equation. The latter  leads to differential equations involving fractional derivatives, either of the Caputo or the Riemann-Liouville  senses, which are equivalent to the special case of the fractional Fokker-Planck equation satisfied by the Mittag-Leffler function known to describe the Cole-Cole relaxation in the time domain. 
\end{abstract}

\keywords{Non-Debye relaxations, Cole-Cole relaxation, Mittag-Leffler function, fractional evolution equations}

\pacs{..}

\maketitle

\section{Introduction}

The Cole-Cole (CC) relaxation model was introduced into dielectric physics by the Cole brothers \cite{KSCole41} to fit experimental data obtained in the measurements of frequency dependence of the electric permittivity. The CC model provides us with an example of non-Debye relaxation for which the spectral function (frequency dependent normalized complex dielectric permittivity) is phenomenologically adjusted to  
\begin{equation}\label{cc}
 \frac{{\hat\epsilon}(\omega) - \epsilon_{\infty}}{\epsilon - \epsilon_{\infty}} = [1 + (\I\!\omega \tau_{0})^{\alpha}]^{-1}. 
 \end{equation}
In the above $\omega$ denotes the frequency, ${\hat\epsilon}(\omega)$ and $\epsilon$ are frequency dependent and static permittivities, respectively, while $\epsilon_{\infty}=\lim\limits_{\omega\to \infty}{\hat\epsilon}(\omega)$ is dielectric constant of induced polarization. Parameters appearing in the RHS {of Eq. \eqref{cc}} come from purely phenomenological analysis:  $\alpha$ is called the width parameter and ranges in the interval $0 < \alpha < 1$; $\tau_{0}$ means an effective time constant related to the so-called loss-peak frequency \cite{CJFBottcher78}. The formalism of the relaxation phenomena theory, namely  the rules which connect the frequency and time regimes, relates the spectral function to the time dependent pulse-response function $f(t/\tau_{0})$ through the Laplace transform ${\cal L}[g(x);s]=\int_{0}^{\infty}e^{-sx}g(x){\rm d}x$  
 \begin{equation}\label{rf}
  \frac{{\hat\epsilon}(\omega) - \epsilon_{\infty}}{\epsilon - \epsilon_{\infty}} = {\cal L}[f(t/\tau_{0}),\I\omega].  
\end{equation} 
Inversion of the Laplace transform \eqref{rf} with the relation \eqref{cc} inserted in is long-time known \cite{Feller66} 
\begin{multline}\label{mlf1}
f(t/\tau_{0}) = \left(\!\frac{t}{\tau_{0}}\!\right)^{-1}\sum\limits_{n=0}^{\infty}\frac{(-1)^{n}(t/\tau_{0})^{\alpha n}}{\Gamma(\alpha n + 1)}\\
= - \frac{{\rm d}}{{\rm d}(t/\tau_{0})} E_{\alpha}[-(t/\tau_{0})^{\alpha}]
\end{multline}
where   $E_{\alpha}(x)$ is the Mittag-Leffler {(ML)} function 
\begin{equation}\label{mlf2}
E_{\alpha}(x)=\sum\limits_{n=0}^{\infty}\frac{x^{ n}}{\Gamma(\alpha n + 1)}
\end{equation}
which {properties have been examined for many years}. The ML function itself, as well as its generalizations, are widely used in many branches of mathematical analysis, first of all in fractional calculus and special functions theory \cite{HJHaubold11,Gorenflo14} {and also in the probability theory, \cite{Korolev172}}.
   
The CC model, being among the oldest and the simplest examples of the non-Debye relaxation, frequently {fits the experimental data of relaxation measurements unsatisfactorily}. In contemporary experimental research it needs to be replaced by more sophisticated models for which it remains a particular case \cite{VVUchaikin13}. Nevertheless, being a training ground of various research concepts it still attracts theoreticians. Their efforts, rooted in the search of physical background of Jonscher's universal relaxation law \cite{Jonscher83,Jonscher96}, are two-fold. The first approach is based on the analysis of stochastic processes supposed to underlie the relaxation phenomena \cite{Weron96,Weron00, Stanislavsky17} and after that the extensive use of generalized central limit theorems \cite{Korolev172, Jurlewicz03, Korolev171}. The alternative method starts from analytical properties of the phenomenologically determined spectral functions. This leads, using tools of the theory of completely monotonic functions \cite{Widder46, Schilling12}, to the time dependent relaxation functions uniquely determined as weighted sums of elementary Debye relaxations \cite{Capelas11, Garrappa16, KGorska18}. It should be noted here that {in} both approaches various Mittag-Leffler type functions appear and do play a very important, even crucial, role.  

Despite of the limited applicability of the CC model in dielectric physics its usefulness goes beyond this branch of physically oriented research and concentrates on various aspects of material science. Examples of its geophysical applications are exhibited, e.g., in \cite{GShen17, ASAhmed17} in which the authors have used it to describe the induced polarization of porous rocks. Applications oriented to the life sciences may be found, e.g., in \cite{HYYe17}  where the CC model has been employed to investigate processes of molecular recognition and also in studies how the age-dependent dielectric properties impact on the brain tissues and proportions of the skull \cite{AChrist10}. Another examples are provided by the  analysis of electric conductivity measured in tissues of the hepatic tumours \cite{DHaemmerich03} and fitting the CC parameters to dielectric data measured in biological tissues and organs \cite{KSasaki14}. A little striking are recently presented applications of the CC model in winery \cite{Machado}. 

In what follows we will adopt a notation $[n(t)/n_0]_{\alpha}$ treated as one symbol parametrized by a real number $\alpha$ and providing us with the ratio of the number $n(t)$ of some objects counted at an instant of time $t$ and the number $n(t_0)\okr n_0$ of the same kind of objects counted at $t_0\le t$. In relaxation processes $[n(t)/n_0]_{\alpha}$ means the relative number (i.e. calculated with respect to the initial number $n_{0}$) of objects which decay, e.g. depolarize, during the time interval $[t_0, t]$. It bears the name of the relaxation function and is defined as a minus primitive of  $f(t/\tau_{0})$, i.e., $\frac{{\rm d}}{{\rm d}(t/\tau_{0})}[n(t)/n_0]_{\alpha}=-f(t/\tau_{0})$. Thus we get
\begin{equation}\label{5/01-0}
\biggl[\frac{n(t)}{n_0}\biggr]_{\alpha}=E_{\alpha}\big[\!-\big(\ulamek{t-t_{0}}{\tau_{0}}\big)^{\!\alpha}\,\big]=E_{\alpha}[\,-(T-T_{0})^{\!\alpha}\,],
\end{equation} 
where, and in what follows, $T_{(\cdot)}$'s denote dimensionless variables  $T_{(\cdot)}=t_{(\cdot)}/\tau_{0}$. It should be also recalled that the function $1-[n(t)/n_0]_{\alpha}$ counts objects which survive the decay in the period {$[t_{0}, t]$} \cite{Weron96}. Obviously, for $\alpha = 1$ the relaxation function $[n(t)/n_0]_{\alpha}$ reduces to the exponential function {$\exp[-(t-t_{0})/\tau_{0}]$} describing the Debye relaxation.

Recall that the basic property of the exponential function is that of being the only solution to the functional equation $g({x+y})=g({x})g({y})$. {This implies that in the Debye case the relaxation function $[n(t_2)/n_0]_{1}$ satisfies the composition law}
\begin{equation}\label{12/03-1}
\left[\frac{n(t_{2})}{n_{0}}\right]_{1} = \left[\frac{n(t_{2})}{n(t_{1})}\right]_{1} \cdot \left[\frac{n(t_{1})}{n_{0}}\right]_{1}
\end{equation}
where the symbol '$\cdot$' denotes the usual multiplication. {An "intermediate"} instant of time $t_{1}\in [t_{0}, t_{2}]$ splits the time interval $[t_{2} - t_{0}]$ into $[t_{2} - t_{1}]$ and $[t_{1} - t_{0}]$. The Debye relaxations are evolution processes without memory \cite{VVUchaikin13,RMHill85}. {For} other evolution patterns, in particular involving the memory effects like it happens in the case of non-Debye models, the composition law given in the form of Eq. \eqref{12/03-1} must not be valid any longer. Describing any non-Debye relaxations we have to adopt another (different from usual multiplication) composition rule if want to describe correctly the evolution $t_{0}\to t_{1}\to t_{2}$. Any proper composition rule must take into account the basic requirement that the evolution in $[t_{0},t_{2}]$ {(and also its functional form)} must coincide with the evolution which starts from some initial condition, goes on to the "intermediate state" (taken at the instant of time $t_{i}$ freely chosen at $[t_{0}, t_{2}]$) and next evolves in $[t_{i},t_{2}]$ reaching the final condition at $t_2$. Illustrative example how to realize such a rule is provided by relaxation described in terms of the stretched exponential functions (the Kohlrausch-Williams-Watts (KWW) ones) for which the composition law is given by the Laplace convolution \cite{KGorska17}. It is natural and absolutely justified to ask for the analogous rule to be obeyed by the CC relaxation. Looking for and finding suitable composition laws satisfied by any model of the time evolution is necessary (although not sufficient) condition for verification its consistency. Such a check is particularly meaningful when one investigates  evolution problems coming out from phenomenology with not fully understood origin in fundamental physics.  The aim of our paper is to fill this gap for the CC model and to present how realization {of} this basic condition, in what follows called {\em evolution consistency requirement}, is achieved in the framework of the CC model. 

The paper is organized as follows. In Sec. \ref{sec2} we recall our main postulate  that the composition of two CC relaxation functions in the time domain leads to another CC relaxation function in the time domain and next we define the analytical form of such a  composition. Its explicit form, if applied in the {\em evolution consistency requirement}, implies the form of the integral evolution equation which the CC relaxation function holds. In Sec. \ref{sec3}, using the numerical calculation we  show that our integral version of the evolution equation is equivalent to its differential version. Differential form of the evolution equation is the fractional Fokker-Planck equation, solvable in terms of  the ML function. The solution of this equation is also given. The similarity between integral and differential forms of the evolution equation shows that it is the fractional kinetics which underlines the CC relaxation. The paper is concluded in Sec. \ref{sec4}.

\section{"Convolution" of the CC relaxations}\label{sec2}

\begin{figure}[!h]
\includegraphics[scale=0.4]{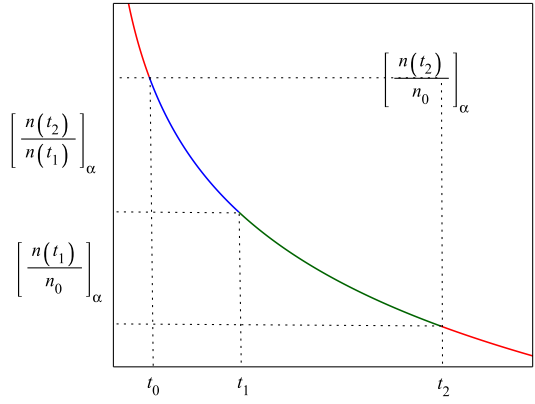}
\caption{\label{fig1}(Color online) Plot of Eq. \eqref{7/01-1} for $\alpha = 1/4$ and $\tau_{0} = 1$. The blue curve is the CC relaxation for $t_{2} - t_{1}$ whereas the green one is for  $t_{1} - t_{0}$. Joining both curves we get the CC relaxation at $t_{2} - t_{0}$ which shape can not depend on the choice of $t_{1}$ for fixed $n(t_{0})$ and $n(t_{2})$.} 
\end{figure}

Let us consider a non-Markovian process for which the composition rule is written down as
\begin{equation}\label{7/01-1}
\left[\frac{n(t_{2})}{n(t_{1})}\right]_{\alpha} \!\circ\! \left[\frac{n(t_{1})}{n(t_{0})}\right]_{\alpha} \!=\! \left[\frac{n(t_{2})}{n(t_{0})}\right]_{\alpha}
\end{equation}
valid for any $t_1$ such that $t_{0} \leq t_{1} \leq t_{2}$. The basic property of the "abstract" composition rule '$\circ$', namely its invariance with respect  to the "intermediate" time $t_{1}$, is graphically illustrated in Fig. \ref{fig1}.

The LHS of Eq. \eqref{7/01-1}, if rewritten in terms of the series representation (cf. Eq. \eqref{mlf2}), gives a double infinite sum. According to the so-called splitting formula for series it may be replaced as
\begin{multline}\label{7/01-2}
\text{LHS of Eq. \eqref{7/01-1}} \\ = \sum_{n=0}^{\infty}(-1)^{n} \sum_{r=0}^{n} \frac{(T_{2}\!-\!T_{1})^{\alpha r} \circ (T_{1}\!-\!T_{0})^{\alpha(n-r)}}{\Gamma(1+\alpha r)\Gamma[1+\alpha(n-r)]}, 
\end{multline}
where, as previously noted,  $T_{i} = t_{i}/\tau_{0}$, $i=0, 1, 2$ are dimensionless variables. Now let us define the composition rule '$\circ$' as the integral
\begin{align}\label{7/01-3}
\begin{split}
(T_{2}-T_{1})^{\alpha r} \circ (T_{1}-T_{0})^{\alpha(n-r)} \okr \frac{1+\alpha n}{1+n} \frac{1}{T_{2} - T_{0}} \\
\times \int_{T_{0}}^{T_{2}}\!\! \D T_{1} (T_{2} - T_{1})^{\alpha r} (T_{1} - T_{0})^{\alpha(n-r)}. 
\end{split}
\end{align}
which explicit evaluation gives
\begin{align}\label{7/01-3a}
\begin{split}
(T_{2}-&T_{1})^{\alpha r} \circ (T_{1}-T_{0})^{\alpha(n-r)}\\
&=\frac{\Gamma(1+\alpha r) \Gamma[1 + \alpha(n-r)]}{(1+n) \Gamma(1+\alpha n)} (T_{2} - T_{0})^{\alpha n}.
\end{split}
\end{align} 
Inserting Eq.\eqref{7/01-3a} into the RHS of Eq. \eqref{7/01-2} and calculating the sums (the sum over $r=0, 1, \ldots, n$ gives $n+1$) we reconstruct Eq. \eqref{7/01-1}. 

To justify the proposed definition of the composition rule '$\circ$' we shall reconsider Eq. \eqref{7/01-2}.  To do this let us look at it from another point of view. Once the RHS of Eq. \eqref{7/01-3} is substituted into Eq. \eqref{7/01-2} we can interchange the order of summations and integration. In the consequence we can write 
\begin{multline}\label{12/01-1}
\text{LHS of Eq. \eqref{7/01-1}} = \int_{T_{0}}^{T_{2}}\!\! \frac{\D T_{1}}{T_{2} - T_{0}} \sum_{n=0}^{\infty} \frac{(-1)^{n}(1+\alpha n)}{(1+n)} \\
\times \sum_{r=0}^{n} \frac{ (T_{2} - T_{1})^{\alpha r} (T_{1} - T_{0})^{\alpha (n-r)}}{\Gamma(1+\alpha r) \Gamma[1 + \alpha(n-r)]}.
\end{multline} 
\noindent {The identification of sums in Eq. \eqref{12/01-1} as coming out from the product of two CC relaxation functions is obstructed by factors  $(1+n)^{-1}$ and $(1+\alpha n)$.} Thus, as expected,  we see explicitly that the operation '$\circ$' must not be interpreted as multiplication. Calculation which shows how to overcome this difficulty and how to represent '$\circ$' by standard mathematical operations is rather laborious and technical so we move it to the Appendix. Final result presented there enables us to rewrite Eq. \eqref{7/01-1}  as
\begin{multline}\label{13/01-2}
\frac{\D}{\D\, t_{2}} \int_{t_{0}}^{t_{2}}\!\! \D t_{1} \int_{0}^{1}\!\! \D u \left[\!\frac{n(u^{\frac{1}{\alpha}} t_{2})}{n(u^{\frac{1}{\alpha}} t_{1})}\!\right]_{\!\alpha}\!\!\!\cdot\left[\!\frac{n(u^{\frac{1}{\alpha}} t_{1})}{n_{0}}\!\right]_{\!\alpha}\!\!= \left[\frac{n(t_{2})}{n_{0}}\right]_{\alpha}\!\!,
\end{multline} 
where the singled out symbol '$\cdot$' denotes usual multiplication. Thus, Eq. \eqref{13/01-2} means that the Eq. \eqref{12/03-1}, expressing "abstract" composition of two CC relaxations, has been represented as the operation which involves appropriately chosen derivative and integral of the usual product. Consequently, we claim that Eq. \eqref{13/01-2}, as  derived from the evolution law Eq. \eqref{7/01-1}, may be treated as the integral form of the evolution equation. 

Validity of Eqs. \eqref{13/01-2} with \eqref{5/01-0} inserted in is checked below for the Debye relaxation, $\alpha = 1$, {case \bf (A)},  
and for the CC relaxation, $0 < \alpha < 1$, {case \bf (B)}. \\

\noindent
{\bf (A)} For the Debye case the LHS of Eq. \eqref{13/01-2} gives
\begin{align}\label{16/01-1}
\begin{split}
&\text{LHS of Eq. \eqref{13/01-2}} = \frac{\D}{\D T_{2}} \int_{T_{0}}^{T_{2}} \D T_{1}\int_{0}^{1}\D u \E^{-u(T_{2}-T_{0})}\\
&\qquad\quad = \frac{\D}{\D T_{2}} \frac{1-\exp[-(T_{2}-T_{0})]}{T_{2} - T_{0}}\int_{T_{0}}^{T_{2}} \D T_{1}\\
&\qquad\quad = \exp[-(T_{2}-T_{0})]=\text{RHS of Eq. \eqref{13/01-2}}.
\end{split}
\end{align}
\\
\noindent
{\bf (B)} For the CC case,  $0 < \alpha < 1$,  we begin with changing the order of integrations in Eq. \eqref{13/01-2} and as the first integral evaluate this over $T_{1}$. Due to Theorem 11.2 of \cite{HJHaubold11}, taken for $\beta=\gamma=\nu=\sigma=1$, we have
\begin{equation}\label{17/01-1}
\int_{0}^{T}\!\! \D\tau E_{\alpha}[-u(T-\tau)^{\alpha}] E_{\alpha}(-u \tau^{\alpha}) = T E_{\alpha, 2}^{2}(-uT^{\alpha})
\end{equation}
where $T = T_{2} - T_{0}$ and $\tau = T_{1} - T_0$. The function $E_{\alpha, \beta}^{\gamma}(\sigma)$, $\sigma\in\mathbb{R}$, $0 < \alpha < 1$, and $\beta, \gamma > 0$, denotes the three parameter generalization of the  ML function $E_{\alpha,\beta}^{\gamma}(\sigma)$. It is defined by the series $E_{\alpha, \beta}^{\gamma}(\sigma)=\sum_{r=0}^{\infty} (\gamma)_{r} \sigma^{r}/[r!\, \Gamma(\beta + \alpha r)]$ with $(\gamma)_{r} = \Gamma(\gamma + r)/\Gamma(\gamma)$ \cite{Garrappa16, HJHaubold11}. The Eq. (11.5) of \cite{HJHaubold11}, used for $\beta = \gamma = 2$, yields
\begin{equation}\label{28/05-1}
\frac{\D}{\D T}T E_{\alpha, 2}^{2}(-uT^{\alpha})=E_{\alpha, 2}^{1}(-uT^{\alpha}).
\end{equation}
The last operation to be done is to integrate Eq. \eqref{28/05-1} over $u\in[0,1]$. It gives $E_{\alpha}(-uT^{\alpha})$, i.e. the RHS of Eq.~\eqref{13/01-2} which ends the proof. 

Here, we would like to remark that Eq. \eqref{13/01-2} is satisfied also for $\alpha > 1$. As an example we consider the ML function for $\alpha = 2$ which is equal to $\cos(T)$. Using it in the Eq. \eqref{13/01-2} gives $E_{2}[-u(T_{2}-T_{1})^{2}]E_{2}[-u(T_{1}-T_{0})^{2}] = \frac{1}{2}\cos[(T_{2}-T_{0})\sqrt{u}] + \frac{1}{2}\cos[(T_{2}-T_{0}-2T_{1})\sqrt{u}]$ which integrated over $T_{1}$, and subsequently differentiated with respect to  $T_{1}$ leads to $\cos[(T_{2}-T_{0})\sqrt{u}] - (T_{2} - T_{0})\sqrt{u} \sin[(T_{2}-T_{0})\sqrt{u}]/2$. Integrating the latter expression over $u$ we reconstruct $E_{2}[-u(T_{2}-T_{0})^{2}]$. 

\section{The differential evolution equation}\label{sec3}

Eq. \eqref{13/01-2} represents the basic evolution law as the integro-differential relation derived from properties of the ML function and the composition rule Eq. \eqref{7/01-3} being assumed or, one may say, even guessed. From our derivation is clear that the ML function satisfies Eq. \eqref{13/01-2} but if we assume the latter as primarily given then we should ask for its solutions. Looking for them we will proceed analogously to the case {\bf B} of the previous section. Evaluating in  Eq. \eqref{13/01-2} the derivative over $t_2$ we get
\begin{align}\label{29/05-1}
\begin{split}
&\frac{\D}{\D\, t_{2}} \int_{t_{0}}^{t_{2}}\!\! \D t_{1} \int_{0}^{1}\!\! \D u \left[\!\frac{n(u^{\frac{1}{\alpha}} t_{2})}{n(u^{\frac{1}{\alpha}} t_{1})}\!\right]_{\!\alpha}\!\!\!\cdot\left[\!\frac{n(u^{\frac{1}{\alpha}} t_{1})}{n_{0}}\!\right]_{\!\alpha}\!\! \\
& \qquad =  \int_{t_{0}}^{t_{2}}\!\! \D t_{1} \int_{0}^{1}\!\! \D u \frac{\D}{\D\, t_{2}}\left[\!\frac{n(u^{\frac{1}{\alpha}} t_{2})}{n(u^{\frac{1}{\alpha}} t_{1})}\!\right]_{\!\alpha}\!\!\!\cdot\left[\!\frac{n(u^{\frac{1}{\alpha}} t_{1})}{n_{0}}\!\right]_{\!\alpha} \\
&\qquad + \int_{0}^{1}\D{u}\left[\!\frac{n(u^{\frac{1}{\alpha}} t_{1})}{n(u_{0})}\!\right]_{\!\alpha} 
= \left[\!\frac{n(u^{\frac{1}{\alpha}} t_{2})}{n_{0}}\!\right]_{\!\alpha}\!\!.
\end{split}
\end{align} 
From the other side, differentiating Eq. \eqref{17/01-1} with respect to $T$, employing \eqref{28/05-1} and integrating the result with respect to $u\in[0,1]$ we arrive at
\begin{align}\label{29/05-2}
\begin{split}
\int_{0}^{1}\D{u}\int_{0}^{T}\D{\tau}\; & \frac{\D}{\D{T}}E_{\alpha}[-u(T-\tau)^{\alpha}]\, E_{\alpha}(-u \tau^{\alpha}) \\
& + \int_{0}^{1}\D{u}E_{\alpha}(-uT^{\alpha}) = E_{\alpha}(-uT^{\alpha}).
\end{split}
\end{align}
Comparison of Eqs. \eqref{29/05-1} and \eqref{29/05-2}, being the same integro-differential relations, leads to the conclusion that
\begin{equation}\label{29/05-3}
\left[\!\frac{n(u^{\frac{1}{\alpha}} t_{2})}{n_{0}}\!\right]_{\!\alpha} = E_{\alpha}[-u(T_{2}-T_{0})^{\alpha}]
\end{equation}
holds at least in a weak sense.

\begin{figure}[!h]
\includegraphics[scale=0.4]{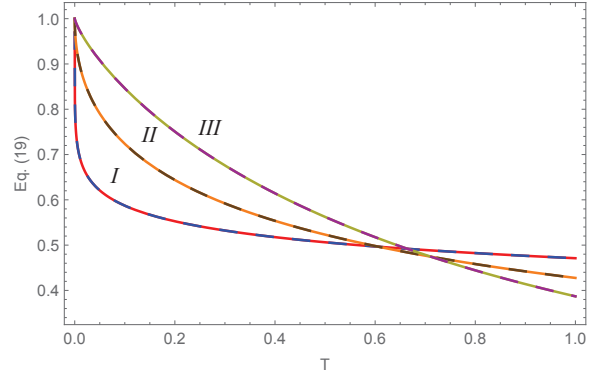}
\caption{\label{fig2}(Color online) Comparison between the LHS of Eq. \eqref{19/01-1} (solid line) and the RHS of Eq. \eqref{19/01-1} (dashed line). The functions are evaluated for $\alpha = 0.2$ (lines no. I), $\alpha = 0.5$ (lines no. II), and $\alpha = 0.8$ (lines no. III).}
\end{figure}

Using the series representation of the two parameter generalized ML function $E^{1}_{\alpha,\beta}(x)$ (special case of that quoted in the previous section, the case {\bf B}) it can be shown that
\begin{equation}\label{19/01-1}
- \int_{0}^{T}\frac{\D\tau}{\Gamma(1-\alpha)}  \frac{E_{\alpha, 0}(-\tau^{\alpha})}{\tau (T-\tau)^{\alpha}} = E_{\alpha}(-T^{\alpha}),
\end{equation}
with $\tau = T_{1}-T_{0}$ and $T = T_{2} - T_{0}$. Eq.\eqref{19/01-1} is illustrated numerically in Fig. \ref{fig2}. 
Employing the Eq. (11.10)  of \cite{HJHaubold11} for $\beta = 1$ leads to
\begin{equation}
\label{28/05-4}
\frac{1}{\tau} E_{\alpha, 0}(-\tau^{\alpha})=\frac{\D}{\D \tau} E_{\alpha}(-\tau^{\alpha})
\end{equation} 
which enables us to rewrite LHS of the Eq. \eqref{19/01-1} as the fractional derivative and Eq. \eqref{19/01-1} as the fractional differential equation. With the substitution of the Eq. \eqref{5/01-0} it yields\begin{equation}\label{4/03-2}
{^{C}\!\partial_{t}^{\alpha}}\!\left[\!\frac{n(t)}{n_{0}}\!\right]_{\alpha} = - \left[\!\frac{n(t)}{n_{0}}\!\right]_{\alpha}, \quad 0 < \alpha < 1,
\end{equation}
with the fractional derivative taken in the Caputo sense, i.e. defined as ${^{C}\partial_{t}^{\alpha}} f(t) = \int_{0}^{t} (t-\tau)^{-\alpha}\frac{\D}{\D\tau} f(\tau) \D\tau/\Gamma(1-\alpha)$ \cite{IPodlubny99}. The relation between the fractional derivatives in the Caputo and the Riemann-Liouvile senses \cite{IPodlubny99} we get the familiar fractional differential equation obeyed by the ML function, see Eq. (10.12) of \cite{HJHaubold11}, for $0 < \alpha < 1$ 
\begin{equation}\label{19/01-2}
{^{RL}\partial_{t}^{\alpha}}\!\left[\!\frac{n(t)}{n_{0}}\!\right]_{\alpha} = - \left[\!\frac{n(t)}{n_{0}}\!\right]_{\alpha} + \frac{t^{-\alpha}}{\Gamma(1-\alpha)}.
\end{equation}
Recall that the fractional derivative in the Riemann-Liouville sense for $0 < \alpha < 1$ is given by ${^{RL}\partial_{t}^{\alpha}} f(t) = \frac{\D}{\D t} \int_{0}^{t} (t-\tau)^{-\alpha} f(\tau) \D\tau/\Gamma(1-\alpha)$.

Eq. \eqref{19/01-2} can be interpreted as the fractional Fokker-Planck equation with the constant operator equal to $-1$ \cite{KGorska12, KGorska-arx17a}. According to \cite{KGorska12, KGorska-arx17a} its formal solution can be written in the form
\begin{equation}\label{29/01-1}
\left[\!\frac{n(t)}{n_{0}}\!\right]_{\alpha} = \int_{0}^{\infty}\!\!\D s\, \E^{-s} \kappa_{\alpha}(s, \ulamek{t}{\tau_{0}}) = \int_{0}^{\infty}\!\!\D s\, \E^{-s} \kappa_{\alpha}(s, T). 
\end{equation}
The integral kernel $\kappa_{\alpha}(x, y)$ is given by $y x^{-(1+1/\alpha)} g_{\alpha}(y x^{-1/\alpha})/\alpha$ with $g_{\alpha}(u)$ being the one-sided L\'{e}vy stable distribution. The explicit form of $g_{\alpha}(u)$ is given in \cite{HPollard46, ELukacs70} whereas in \cite{KAPenson10, KGorska12a, KGorska17} it is  represented as finite sum of the generalized hypergeometric functions. The substitution Eq. \eqref{29/01-1} into Eq. \eqref{13/01-2} allows one to obtain
\begin{align}\label{29/01-3}
\begin{split}
\kappa_{\alpha}(x, T) & = \frac{\D}{\D T} \int_{0}^{T} \D\tau \int_{0}^{1} \D u \int_{0}^{x}\D\xi \\
& \times \kappa_{\alpha}[\xi, u^{1/\alpha}(T-\tau)]\; \kappa_{\alpha}(x-\xi, u^{1/\alpha}\tau),
\end{split}
\end{align}
where $T = T_{2} - T_{0}$ and $\tau = T_{1} - T_{0}$. 

\section{Conclusion}\label{sec4}

We have studied the consequences of the physically natural composition rule required to be obeyed by the time evolution of the CC relaxation function. Condition which we have been proposing reflects the rule that the CC relaxation taking place at the time interval $t_{2} - t_{0}$ can be composed from CC relaxations at the times $t_{2} - t_{1}$ and $t_{1} - t_{0}$, where $t_{0} \leq t_{1} \leq t_{2}$. If such a composition rule is represented as the multiplication then we deal with the standard Debye relaxation which may be interpreted as the special case of CC relaxation for the width parameter $\alpha = 1$. In the general case of the CC relaxation, $\alpha \neq 1$, we have shown that the composition of two CC relaxations functions must not simplify to  multiplication and has to be understood as a combination of integration and differentiation of their product, see Eq. \eqref{13/01-2}. The explicit form of the composition law provides us with the integro-differential relation which we identify as the integral evolution equation governing the CC relaxation. Next, we have rewritten this equation as the fractional differential equation with the fractional derivatives taken either in the Riemann-Liouville or the Caputo sense. As should be expected these are the equations known to be satisfied by the ML function. Moreover, formal solutions to these equations are given via the well-known integral formulae relating, through the Laplace-like transforms, the ML function with the integral kernel function involving the one-sided L\'{e}vy stable distribution. 

Resuming our work we also would like to point out that the integral evolution equation Eq. \eqref{13/01-2} exhibits not only the time evolution of the CC relaxation function but also the composition properties of the ML functions which is the problem interesting mathematically at its own. The ML function is the eigenfunction of the standard fractional derivative, namely the Caputo one, and in the fractional calculus plays the same role as the exponential function does in conventional calculus. This is particularly well seen when one investigates methods of solving fractional differential equations and explains why the ML function is so useful and has many applications in fractional dynamics. As a guiding example we mention the anomalous diffusion \cite{RMetzler99, RMetzler00, IMSokolov05, JMSancho04} described by the ML function which emerges as the solution of the fractional Fokker-Planck equation governing the process \cite{KGorska12, KGorska-arx17a, EBarkai01}. Besides of knowledge of the time evolution equation and its solution the knowledge and verification of the composition law it satisfies remains the universal consistency check of the results obtained within proposed approach. We consider our results as a significant step in this direction. 

\appendix
\section*{Appendix}
In the formula Eq.\eqref{12/01-1}, namely 
\begin{multline}
\text{LHS of Eq. \eqref{7/01-1}} = \int_{T_{0}}^{T_{2}}\!\! \frac{\D T_{1}}{T_{2} - T_{0}} \sum_{n=0}^{\infty} \frac{(-1)^{n}(1+\alpha n)}{(1+n)} \nonumber\\
\times \sum_{r=0}^{n} \frac{ (T_{2} - T_{1})^{\alpha r} (T_{1} - T_{0})^{\alpha (n-r)}}{\Gamma(1+\alpha r) \Gamma[1 + \alpha(n-r)]}\nonumber
\end{multline}
we: 
\begin{itemize}
\renewcommand{\labelitemi}{i)}
\item{replace $(1+n)^{-1}$ by the integral  $\int_{0}^{1} u^{n} \D u$,}
\renewcommand{\labelitemi}{ii)}
\item{instead of $(1+\alpha n) {(T_{2} - T_{1})}^{\alpha r} {(T_{1} - T_{0})}^{\alpha (n-r)}$  write 
\begin{align*}
&\quad{(T_{1} - T_{0})^{\alpha(n-r)}(T_{2} - T_{1})^{\alpha r}}\\ 
&\qquad {+ (T_{2} - T_{1})^{\alpha r} (T_{1} - T_{0})\D_{\; (T_{1} - T_{0})} (T_{1} - T_{0})^{\alpha (n-r)}}\\
&\qquad {+(T_{1} - T_{0})^{\alpha(n-r)} (T_{2} - T_{1})\D_{\;(T_{2} - T_{1})} (T_{2} - T_{1})^{\alpha r}}, 
\end{align*}
where the abbreviation  $\D_{\;a} f(a)$ is used for $\D f(a)/\D\, a$.}
\end{itemize}
Inserting these expressions into Eq. \eqref{12/01-1} and using Eq. \eqref{mlf2} we arrive at
\begin{align}\label{12/01-1a}
\begin{split}
& \text{LHS of Eq. \eqref{7/01-1}} \!=\!\! \int_{0}^{1}\!\!\!\frac{\D u}{T_{2}\! -\!T_{0}} \int_{T_{0}}^{T_{2}}\!\!\! \D T_{1}\\ 
&\qquad \times \biggl\{E_{\alpha}[-u(T_{2} - T_{1})^{\alpha}]  E_{\alpha}[-u(T_{1}-T_{0})^{\alpha}] \biggr. \\
&\qquad + (T_{1}-T_{0}) E_{\alpha}[-u(T_{2}-T_{1})^{\alpha}] \frac{\D E_{\alpha}[-u(T_{1} - T_{0})^{\alpha}]}{\D\; (T_{1}-T_{0})} \\
&\qquad \left.+ (T_{2}-T_{1}) E_{\alpha}[-u(T_{1}-T_{0})^{\alpha}]  \frac{\D E_{\alpha}[-u(T_{2} - T_{1})^{\alpha}]}{\D\; (T_{2}-T_{1})}\right\}.
\end{split}
\end{align}
To simplify the above let us: 
\begin{itemize}
\renewcommand{\labelitemi}{iii)}
\item{replace $\D_{\;(T_{1} - T_{0})}$ by $\D_{\;T_{1}}$ and $\D_{\;(T_{2} - T_{1})}$ by $-\D_{\;T_{1}}$;}
\renewcommand{\labelitemi}{iv)}
\item{apply in Eq. \eqref{12/01-1a} the Leibniz  rule to one of the products containing the ML function and its derivative, e.g. to that in the third line} 
\begin{align*}
&E_{\alpha}[-u(T_{2}-T_{1})^{\alpha}] \D_{\; (T_{1}-T_{0})} E_{\alpha}[-u(T_{1}-T_{0})^{\alpha}]\\
&\qquad\pokr\D_{\;T_{1}}\bigl\{E_{\alpha}[-u(T_{2}-T_{1})^{\alpha}] E_{\alpha}[-u(T_{1}-T_{0})^{\alpha}]\bigr\} \\
&\qquad-E_{\alpha}[-u(T_{1}-T_{0})^{\alpha}] \D_{\;T_{1}}E_{\alpha}[-u(T_{2}-T_{1})^{\alpha}]\\
&\qquad\pokr\D_{\;T_{1}}\bigl\{E_{\alpha}[-u(T_{2}-T_{1})^{\alpha}] E_{\alpha}[-u(T_{1}-T_{0})^{\alpha}]\bigr\}\\
&\qquad+E_{\alpha}[-u(T_{1}-T_{0})^{\alpha}] \D_{\;T_{2}}E_{\alpha}[-u(T_{2}-T_{1})^{\alpha}].
\end{align*}
\end{itemize}
With these substitutions Eq. \eqref{12/01-1a} becomes
\begin{align}\label{12/03-2}
\begin{split}
& \text{LHS of Eq. \eqref{7/01-1}} = \int_{0}^{1}\!\!\D u \int_{T_{0}}^{T_{2}}\!\!\frac{\D T_{1}}{T_{2} - T_{0}} \\
& \times \biggl\{E_{\alpha}[-u(T_{2}-T_{1})^{\alpha}] E_{\alpha}[-u(T_{1}-T_{0})^{\alpha}]\biggr. \\
& +\biggl.(T_{1}-T_{0}) {\D_{\; T_{1}}}\bigl\{E_{\alpha}[-u(T_{2}-T_{1})^{\alpha}] E_{\alpha}[-u(T_{1}-T_{0})^{\alpha}]\bigr\}\biggr\} \\
& - \int_{0}^{1}\!\!\D u \int_{T_{0}}^{T_{2}}\!\! \D T_{1} E_{\alpha}[-u(T_{1} - T_{0})^{\alpha}] {\D_{\; T_{1}}} E_{\alpha}[-u(T_{2}-T_{1})^{\alpha}]. 
\end{split}
\end{align}
Employing the Leibniz rule to the expression in the third line of Eq. \eqref{12/03-2} implies that the first term in the curly brackets cancels and the evaluation of the first integral in Eq. \eqref{12/03-2} over $T_1$ gives   
\begin{align*}
&(T_{2}-T_{0})^{-1}\int_{0}^{1}\D u \biggl\{(T_{1} - T_{0}) E_{\alpha}[-u(T_{2}-T_{1})^{\alpha}] \biggr.
\\
&\biggl.\times E_{\alpha}[-u(T_{1}-T_{0})^{\alpha}]\biggr\}\big\vert_{T_{1}=T_{0}}^{T_{1}=T_{2}}=\int_{0}^{1} \D u E_{\alpha}[-u(T_{2}-T_{0})^{\alpha}],
\end{align*}
where it has been employed that the ML function equals 1 for vanishing argument. Taking this result into account and changing in the second integral in  Eq. \eqref{12/03-2} the derivative with respect to $T_{1}$ into the derivative with respect to $(T_{2}-T_{1})$ we get that the RHS of the Eq. \eqref{12/03-2} becomes
\begin{multline}\label{13/01-1}
\int_{0}^{1}\!\! \D u \int_{T_{0}}^{T_{2}}\!\! \D T_{1}\, \frac{\D E_{\alpha}[-u(T_{2} -T_{1})^{\alpha}]}{\D\; (T_{2} - T_{1})} E_{\alpha}[-u(T_{1} - T_{0})^{\alpha}] \\
+ \int_{0}^{1}\!\! \D u E_{\alpha}[-u(T_{2} - T_{0})^{\alpha}]. 
\end{multline} 
Moving  the derivative $\D/\!\!\D\,(T_{2} - T_{1})$ in front of the first integral in the Eq. \eqref{13/01-1}
\begin{align}\label{22/05_1}
&\int_{0}^{1}\!\! \D u \int_{T_{0}}^{T_{2}}\!\! \D T_{1}\, \frac{\D E_{\alpha}[-u(T_{2} -T_{1})^{\alpha}]}{\D\; (T_{2} - T_{1})} E_{\alpha}[-u(T_{1} - T_{0})^{\alpha}] \nonumber \\
&\quad = \frac{\D}{\D\; (T_{2} - T_{1})}\int_{0}^{1}\!\! \D u \int_{T_{0}}^{T_{2}}\!\! \D T_{1}\,E_{\alpha}[-u(T_{2} -T_{1})^{\alpha}]\\
&\quad \times E_{\alpha}[-u(T_{1} - T_{0})^{\alpha}] - \int_{0}^{1}\!\! \D u E_{\alpha}[-u(T_{2} - T_{0})^{\alpha}] \nonumber
\end{align}
we end up with the equation
\begin{multline*}
\frac{\D}{\D\; T_{2}}\int_{0}^{1}\!\! \D u \int_{T_{0}}^{T_{2}}\!\! \D T_{1}\,E_{\alpha}[-u(T_{2} -T_{1})^{\alpha}] E_{\alpha}[-u(T_{1} - T_{0})^{\alpha}] \\ 
= E_{\alpha}[-(T_{2} \!-\! T_{0})^{\alpha}], \nonumber
\end{multline*}
{which after using Eq. \eqref{5/01-0} gives Eq. \eqref{13/01-2}.}

\section*{Acknowledgments}
The authors were supported by the NCN research project OPUS 12 no. UMO-2016/23/B/ST3/01714. K. G. acknowledges the support of the MNiSW (Warsaw, Poland) Programme "Iuventus Plus 2015-2016", project no IP2014 013073 and the NCN Programme Miniatura 1, project no. 2017/01/X/ST3/00130.

\end{document}